\documentclass[a4paper,fleqn,usenatbib,useAMS]{mnras}
\usepackage{graphics}
\usepackage{graphicx}
\usepackage[pagewise]{lineno}
\usepackage{amsmath}    
\usepackage{amssymb}    
\usepackage{multicol}        
\usepackage{bm}     
\usepackage{pdflscape}  
\usepackage{xcolor}
\usepackage{ae,aecompl}
\usepackage{subfigure}
\usepackage{scalerel}
\usepackage[english]{babel}
\usepackage{times}
\usepackage{physics}
\usepackage{etoolbox}
\usepackage{multirow, booktabs}
\usepackage{siunitx}
\usepackage{paralist}
\usepackage{makecell}
\usepackage{tabularx}
\usepackage[inline]{enumitem}
\newlist{mycompactenum}{enumerate}{1}
\setlist[mycompactenum,1]{nosep,label=\arabic*.}
\usepackage{paralist}

\newcommand{\wfirst}{\textit{Roman}}

\title[Improved AE algorithm and its application]{Improved Aberth-Ehrlich root-finding algorithm and its further application for Binary Microlensing}
\author[ H. Fatheddin, S. Sajadian]{Hossein Fatheddin$^{1}$\thanks{E-mail: Fathoaldin@ph.iut.ac.ir},
Sedighe Sajadian$^{1}$\thanks{E-mail: s.sajadian@iut.ac.ir}\\
$^{1}$Department~of~Physics,~Isfahan~University~of~Technology,~Isfahan~84156-83111,~Iran}

\begin{document}

\label{firstpage}
\pagerange{\pageref{firstpage}--\pageref{lastpage}}
\maketitle
\begin{abstract}
In gravitational microlensing formalism and for modeling binary light curves, the key step is solving the binary lens equation. Currently, a combination of the Newton's and Laguerre's methods which was first introduced by Skowron \& Gould (SG) is used while modeling binary light curves. In this paper, we first introduce a fast root-finding algorithm for univariate polynomials based on the Aberth-Ehrlich (AE) method which was first developed in 1967 as an improvement over the Newton's method. AE algorithm has proven to be much faster than Newton's, Laguerre's and Durand-Kerner methods and unlike other root-finding algorithms, it is able to produce all the roots simultaneously. After improving the basic AE algorithm and discussing its properties, we will optimize it for solving binary lens equations, which are fifth degree polynomials with complex coefficients. Our method is about $1.8$ to $2.0$ times faster than the SG algorithm. Since, for calculating magnification factors for point-like or finite source stars, it is necessary to solve the binary lens equation and find the positions of the produced images in the image plane first, this new method will improve the speed and accuracy of binary microlensing modeling. \\
\end{abstract}


\section{Introduction}
Studying the effects of binary microlensing has led to important astrophysical applications like detecting extrasolar planets orbiting the lens objects \citep{Bennett1996,Bennett2008,1991MAO,Gould1992,2021sajadian} or even isolated and dark black holes in the Galactic disc \citep{2022Sahu}. The current lensing and microlensing projects, i.e., the Optical Gravitational Lensing Experiment (OGLE, \citet{OGLE_IV}), the Microlensing Observations in Astrophysics (MOA, \citet{MOA_gourp}) microlensing group, and the Korea Microlensing Telescope Network (KMTNet, \citet{KMTNet2016}), produce an increasing amount of data and the future implementation of ground and Space based surveys, like Euclid mission \citep{Penny2013} and \textit{The Nancy Grace Roman Space Telescope} (\wfirst)~survey \citep{Penny2019} will produce even more data for these events. So, the fast and accurate modeling of this information is of great importance.\\

The microlensing events due to single and point lens objects have been known to have symmetric magnification curves. These events have simple lens equation with a light deflection angle that is defined as:
\begin{eqnarray}
\label{eq:deflection}
\alpha(\textbf{x})=\frac{4GM}{c^2}[\frac{(\textbf{x}-\textbf{x}_l)}{(\textbf{x}-\textbf{x}_l)^2}]
\end{eqnarray}
Where G is the universal gravitational constant, c is the speed of light, M is mass of the lens and $\textbf{x}_l$ is its position vector. For a point-like source star, two images are formed which give the magnification factor as \citep[see, e.g.,][]{gaudi2012}:  
\begin{eqnarray}
A= \frac{u^{2}+2}{u \sqrt{u^{2}+4}},~~~ u=\sqrt{u_{0}^{2}+  (\frac{t-t_{0}}{t_{\rm E}})^{2}},  
\end{eqnarray}

\noindent where, $u_{0}$ is the lens impact parameter (with origin on the center of mass), $t_{0}$ is the time of the closest approach, and $t_{\rm E}$ is the Einstein crossing time which represents the time scale of a lensing event. Here, $u$ is the lens-source distance projected on the lens plane and normalized to the Einstein radius, $R_{\rm E}$, which is the radius of the images ring at the time of complete alignment:
$$R_{\rm E}=\sqrt{\frac{4~G~M_{\rm l}}{c^{2}} \frac{D_{\rm l}~D_{\rm{ls}}}{D_{\rm s}}},$$
where, $M_{\rm l}$ is the lens mass,  $D_{\rm{ls}}=D_{\rm s}-D_{\rm l}$, $D_{\rm l}$ and $D_{\rm s}$ are the lens and source distance from the observer. For an extended source star, the magnification factor is expressed in terms of elliptical integrals which is the so-called finite source size effect \citep{1994wittmoa}. This effect is considerable for events with small $R_{\rm E}$ values, such as the events due to free-floating exoplanets \citep{2019Mroz,2021sajadianb}.   

Compared to modeling of simple and symmetric  single-lens microlensing light curves which is very fast, binary microlensing light curves can not be modeled analytically. In order to calculate the magnification factors and light curves in the case of binary lenses, it is important to solve the binary lens equation to find the positions of images for a point-like source star. In this case we need to introduce more parameters which allows for a very large variety of binary lens light curves and makes binary lenses more complicated than a single one. The case of binary lens equation and its solutions is discussed in details in section \ref{sec: binary}.

\noindent After discussing the case of binary lenses, we will introduce a new root-solving algorithm that can solve univariate polynomials with complex or real coefficients with high accuracy in a very short time. This algorithm will be be developed based on the Aberth-Ehrlich method. However, we determine initial guess for the roots by improving a method first introduced by Guggenheimer \citep[see][]{Guggenheimer1986}. We introduce our method with a brief introduction on different varieties of polynomial solving methods, in Section \ref{sec:Aberth}. In the last section, we discuss and summarize the results. 

\section{Binary Lenses}\label{sec: binary}

As mentioned in the introduction, The Binary Lens equation gives rise to a fifth degree polynomial. In this section we are going to review some characteristics of a binary lens event and its equation so that we can develop and optimize our root-finding algorithm in the next section.

\subsection{Properties of a Binary Lens system}
Properties of a system containing two lenses have been studied in great detail before \citep[see][]{1986Schneider}. When a source passes from behind two lenses, the only change from a single lens situation (equation~\ref{eq:deflection}) is that the deflection angle consists of the sum of two point lenses:
\begin{eqnarray}
\label{eq:Lens Eq}
\alpha(\textbf{x})=\frac{4G}{c^2}(\frac{m_1(\textbf{x}-\textbf{x}_1)}{(\textbf{x}-\textbf{x}_1)^2}+\frac{m_2(\textbf{x}-\textbf{x}_2)}{(\textbf{x}-\textbf{x}_2)^2})
\end{eqnarray}
\noindent
here $m_1$, $m_2$ are the masses of the two lenses and $\textbf{x}_1$,$\textbf{x}_2$ are their positions. Since equation~\ref{eq:Lens Eq} is nonlinear, for the case of binary lenses we can not simply sum two single lens cases and the caustics, magnification maps and light curves look very different.

\indent In a binary lens scenario, we use the following parameters to analyze or model an event:
\begin{itemize}
\item the mass ratio $q=m_1/m_2$,
\item the binary separation d (in units of the Einstein radius for the total masses; $m=m_1+m_2$),
\item the angle $\phi$ between the source trajectory and the line connecting the two lenses,
\item Radius of the source $\rho_{\rm *}$ (also in the Einstein Radius),
\item the impact parameter $u_0$,
\item time of the closest approach $t_0$ and the Einstein crossing time $t_{\rm E}$ .
\end{itemize} 
This allows for a very large variety of binary lens light curves and makes binary lenses more complicated than a single one, which was reviewed in the introduction.
\subsection{The Lens Equation}
 The first attempt for solving the binary lens equation (when lenses have equal masses, i.e. q=1) was done by P. Schneider and his collaborators in \citet{1986Schneider,1992schneider, 1993Erdlschneider}. In complex notation, the binary lens equation was derived by \citet{Witt1990}, as given by:
\begin{eqnarray} \label{eq:lenseq1}
\zeta=z+\frac{m_{\rm 1}}{\bar{ z_{\rm 1}}-\bar{z}}+\frac{m_{\rm 2}}{\bar{z_{\rm 2}}-\bar{z}} 
\end{eqnarray}
Where $m_1=1/(1+q)$ and $m_2=q/(1+q)$ are the masses of the lenses normalized to the total mass of lenses. $z_1$ and $z_2$ are the positions of the lenses, $\zeta=\xi+i\eta$ and $z=x+iy$ are the source and the image positions and $\bar{z}$ indicates the complex conjugate of z.  

\noindent By setting both lenses over the $x$-axis symmetrically with respect to the center, i.e., $z_1=-z_2=(-d/2,~0)$, the lens equation \ref{eq:lenseq1} arrives at a lengthy fifth order polynomial in z \citep{WittMao1995}:
\begin{eqnarray}\label{eq:lenseq}
p_{\rm 5}(z)=\sum_{\rm i=1}^{5} c_i z^i=0 ,
\end{eqnarray}
where,
\begin{flushleft}
$c_5 = z_1^2 - \bar{\zeta}^2$,~~~~~~~~~~~$c_4=-2 m \bar{\zeta} ~+~ \zeta \bar{\zeta}^2 ~-~ 2 \Delta m z_1 ~-~ \zeta z_1^2$, \\
$c_3 = 4 m \zeta \bar{\zeta} ~+~ 4 \Delta m \bar{\zeta} z_1 ~+~ 2 \bar{\zeta}^2 z_1^2 ~-~ 2 z_1^4$ ,\\
$c_2 = 4 m^2 \zeta ~+~ 4 m \Delta m z_1 ~-~ 4 \Delta m \zeta \bar{\zeta} z_1 ~-~ 2 \zeta \bar{\zeta}^2 z_1^2$ \\ $~~~~~~~~~~~~+~ 4 \Delta m z_1^3 ~+~ 2 \zeta z_1^4 $ ,\\
$c_1 =  -8 m \Delta m \zeta z_1 ~-~ 4 (\Delta m)^2 z_1^2 ~-~ 4 m^2 z_1^{2} ~-~ 4 m \zeta \bar{\zeta} z_1^2 $\\
$~~~~~~~~~~~~~~~~-~ 4 \Delta m \bar{\zeta} z_1^3 ~-~\bar{\zeta}^2  z_1^4  ~+~ z_1^6,$\\
$c_0 = z_1^2 \Big[4 (\Delta m)^2 \zeta ~+~ 4 m \Delta m z_1 ~+~ 4 \Delta m \zeta \bar{\zeta} z_1 ~+~ 2 m \bar{\zeta} z_1^2 $\\
$~~~~~~~~~~~~~+~ \zeta \bar{\zeta}^2 z_1^2 ~-~ 2 \Delta m z_1^3 -\zeta z_1^4 \Big]$.
\end{flushleft}

\noindent here, $m=(m_{1}+m_{2})/2$ and $\Delta m=(m_{2}-m_{1})/2$. In general, it is not possible to analytically solve this equation. We need a numerical root-solving method that is sufficient, fast and accurate to trace the lens equation.

For the case of Point-like sources ($\rho_{\rm *}=0$), we simply need to trace source trajectory and solve equation~\ref{eq:lenseq} for z, which is in the format of equation~\ref{eq:poly}. So, we can solve it by the root-solving algorithm in section~\ref{sec:Aberth}. Since the source is passing behind the lenses, in order to find positions of the produced images ($z=x+iy$), different iterations are made for each source position ($\zeta=\xi+i\eta$). 
\noindent Additionally, for microlensing light curves of an extended source star the first step is solving the binary lens equation, \ref{eq:lenseq}, for the points over the source circle to find the circumference of images. In the next step, using either the Green's theorem \citep{1997Gould,1998Dominik,Bozza2010}, or inverse ray shooting \citep{1986Kayser,1990Wambsganss,1999Wambsganss}, or a tree algorithm \citep{2010sajadian} the area of the images will be calculated according to their environments, which leads to the magnification factor and light curves. Based on these methods, some python packages were developed up to now \citep[see, e.g., ][]{pyLima, MuLensModel} which were mostly based on a microlensing computation code in C++ described in \citet{2018Bozza}. Hence, their common step is solving the binary lens equation.
  
In the next section, we are going to develop and optimize our root-finding algorithm for solving the special equation~\ref{eq:lenseq}.

\section{Root-Finding Algorithm} \label{sec:Aberth}
The attempts to numerically find roots of a polynomial can be traced back to Newton (1671) and Halley (1694). But the first derivative based iterative root-finding algorithm was first introduced by Thomas Simpson in 1740 and today, it is generally known as the Newton-Raphson method \citep[see:][]{Simpson1740,SkowronGould2012}. Since then, the algorithm has been improved multiple times by various mathematicians. Here, we introduce the root-finding algorithm based on an improvement made by Aberth (1973) and Ehrlich (1967).\\
\noindent The Aberth-Ehrlich (AE) method introduced by \citet{Ehrlich1967,Aberth1973} combines the elements of Newton's method with an implicit deflation strategy, which allows for the computation of all roots of a polynomial simultaneously and converges cubically. This method is considered an improvement over the Durand-Kerner method, another simultaneous root solver method which converges quadratically and is 10-100 times slower \citep[see, e.g., ][]{Ghidouche2017}.

\noindent The facts that AE is extremely fast for various degrees of polynomials, its ability to find all the roots at once (unlike other iterative root-solving methods such as Laguerre's) and its root polishing procedure, which is inside the main iteration loop and can be controlled for a chosen accuracy, suggest a great numerical solution for different polynomials.

\indent We derive and explain our method in the following subsection. Afterwards, we illustrate our method for determining the initial estimation of the roots In the subsection \ref{Initial_guess}. Some features regarding error and the stopping criteria are discussed in subsection \ref{errorstop}. The advantage of this method in the regard of execution time in comparison with other root-finding methods is argued in the last subsection.

\subsection{Aberth-Ehrlich Algorithm and its derivation}\label{sub1} 
\noindent For the derivation of the AE algorithm, we start from Newton's method by considering a univariate $n$th order polynomial of the form:
\begin{eqnarray} \label{eq:poly}
p_{n}(x)=c_n x^n + c_{n-1} x^{n-1} + ... + c_1 x + c_0, 
\end{eqnarray}
where, $c_{i}$ denote real or complex coefficients. Then, since we need $n$ distinct roots for a polynomial of degree $n$, we assume $n$ distinct guesses in the forms of: $\lambda_1^{(0)}$, $\lambda_2^{(0)}$, ..., $\lambda_n^{(0)}$. The superscript (0) denotes that the guesses are primary and have not entered the iteration process yet.
The iteration in Newton's method is:
\begin{eqnarray} \label{eq:Newton}
\lambda_i^{(k+1)}=\lambda_i^{(k)}-\frac{F(\lambda_i^{(k)})}{F'(\lambda_i^{(k)})}
\end{eqnarray}
Where F is a function of polynomial. Now if we consider:
\begin{eqnarray} \label{eq:F1}
F(x)=\frac{p_{n}(x)}{\Pi _{j=1,j\neq k}^n (x-\lambda_j^{(k)})},
\end{eqnarray}
Instead of simply putting $F(x)=P_n(x)$, then:
\begin{flalign*}
\frac{F'(x)}{F(x)}&=\frac{d}{dx}ln \abs{F(x)}
&=\frac{d}{dx} (ln \abs{P_n(x)}-\sum_{\rm j=1, j\neq k}^{n} \abs{(x-\lambda_j^{(k)}})\\
&=\frac{P'_n(x)}{P(x)}-\frac{1}{\sum_{\rm j=1,j\neq k}^n (x-\lambda_j^{(k)})}\\
\end{flalign*}
So, we have:
\begin{eqnarray}
\frac{F(x)}{F'(x)}=\frac{1}{\frac{P'_n(x)}{P(x)}-\frac{1}{\sum_{\rm j=1,j\neq k}^n (x-\lambda_j^{(k)})}}
\end{eqnarray}
If we put this equation in the Newton's iteration (equation \ref{eq:Newton}), we arrive at the AE algorithm: 
\begin{eqnarray} \label{eq:Aberth}
\lambda_i^{(k+1)}=\lambda_i^{(k)}-\frac{\alpha_i^{(k)}}{1-\alpha_i^{(k)}\beta_i^{(k)}},
\end{eqnarray}

\noindent where: 
\begin{eqnarray}
\alpha_i^{(k)}=\frac{p_{n}(\lambda_i^{(k)})}{p'_{n}(\lambda_i^{(k)})} ,~~~~~~~~~~~~~ \beta_i^{(k)}=\sum_{\rm i\neq j}^{n} \frac{1}{\lambda_i^{(k)}-\lambda_j^{(k)}},
\end{eqnarray}
here $p'_{n}(\lambda_i^{(k)})=dp_{n}/dx$ is the derivative of the univariate polynomial $p_{n}$ at $x=\lambda_i^{(k)}$. We note that in the Newton's method the iteration process for finding the roots is:
\begin{eqnarray}
\lambda_i^{(k+1)}=\lambda_i^{(k)}-\alpha_i^{(k)}
\end{eqnarray}
The equation~\ref{eq:Aberth} presents a single Newton iteration on equation~\ref{eq:F1} to determine the root $\lambda_i$.

Conceptually, AE is analogous to an electrostatic case \citep[see: ][]{Aberth1973}, the initial roots ($ \lambda_i^{(0)}$) are modeled as free negative point charges, which converge toward the actual roots that are represented by fixed positive point charges. If the newton's method was used directly, some of the initial roots would incorrectly converge to the same root and $n$ distinct roots could not be found. AE avoids this by  modeling the repulsive force between the movable charges. In this way, when a movable charge has converged to a root, their charges will cancel out, so that other movable charges are no longer attracted to that location, forcing them to converge to other unoccupied roots.

In the next subsection, we explain our method of determining the initial guess for the roots, i.e., $ \lambda_i^{(0)}$.   

\begin{table}
	\centering
	\begin{tabular}{cc}
		\hline
		\textbf{Degree of Polynomial} & \textbf{Number of Iterations} \\
		\hline
		3 & 3 \\
		4 & 4 \\
		5 & 5 \\
		6 & 5 \\
		7 & 5 \\
		8 & 6 \\
		9 & 6 \\
		10 & 7 \\
		50 & 12 \\
		100 & 20\\
		200 & 32\\
		300 & 43\\
		\hline
	\end{tabular}
	\caption{This table gives the number of iterations that are needed for finding all the roots based on the degree of polynomial in our method}
\label{tab:iter}
\end{table}
\subsection{The Initial guess problem} \label{Initial_guess}

The iteration method (Equation~\ref{eq:Aberth}) requires the input of some initial guess to start the iteration process. Numerical experiments show that the number of iterations needed for convergence depends greatly and somewhat erratically on the choice of  $ \lambda_i^{(0)}$. If $ \lambda_i^{(0)}$s are more distant from the actual roots, the number of iterations that are needed for convergence increases and the whole process becomes time-consuming. \citet{Ehrlich1967} recommended a choice of $ \lambda_i^{(0)}$ between the maximum and the minimum of the absolute values of the roots of the equation by defining an Upper bound and a Lower bound for the roots. The condition $ \lambda_i^{(0)}>\lambda_{min}^{(*)}$ (where $\lambda_{min}^{(*)}$ is the minimum root of the polynomial) is natural since in most cases the choice $ \lambda_i^{(0)}<\lambda_{min}^{(*)}$ leads to instability and does not converge. The other condition, $ \lambda_i^{(0)}<\lambda_{max}^{(*)}$, may seem unnecessary but is quite helpful, since it reduces the number of needed iterations.

\noindent \citet{Guggenheimer1986} presents the following method for finding $ \lambda_i^{(0)}$ based on the above idea:

First we calculate $U_i$ and $V_i$, where $U_i$ is an upper bound and $V_i$ is a lower bound for each root, as the following:
\begin{eqnarray}
U_i &=& 2 \abs{ \frac{c_i}{ c_0} } ^{1/i}, \nonumber \\
V_j &=& \frac{1}{2\abs{c_j / c_n}^{\frac{1}{n-j}}},
\end{eqnarray}
\noindent where, $i$ changes from 1 to $n$, whereas $j$ changes from $0$ to $n-1$. Then, after computing the average values $\left<U \right>_i$ and $\left<V \right>_i$, we delete the $U_i$ for which $U_i - \left<U\right>_i$  is maximal and delete the $V_j$ for which $\left<V \right>_i-V_j$ is maximal. Afterwards, we recompute $\left<U\right>_{i+1}$ and $\left<V \right>_{i+1}$. each time, the initial guess in this method would be:
\begin{eqnarray}
\lambda_i^{(0)}=\frac{\left<U\right>_i+\left<V\right>_i}{2}.
\end{eqnarray}
\noindent In this method we have to calculate different values for U and V, and by computing their averages, find the value for $\lambda_i^{(0)}$.

\indent Here, we improve this method by considering $U$ and $V$ ,The upper and lower bounds of all roots (instead of each individual root), as:
\begin{eqnarray}\label{eqqq}
U&=&1+\frac{\max{\big(\abs{c_1},~\abs{c_2},  ... ,~\abs{c_n} \big)} }{\abs{c_n}},\nonumber \\
V&=&\frac{\abs{c_0}}{\abs{c_0}+ \max{\big(\abs{c_1},~\abs{c_2},  ... ,~\abs{c_n}\big)}  },
\end{eqnarray}

\noindent where, $\abs{c_{i}}=\sqrt{c_{i} \bar{c_{i}}}$. Then, we randomly choose $n$ distinct numbers from a normal distribution between these lower and upper values, i.e., $r_{i} \in [U,~V]$. Afterwards, we define:

\begin{eqnarray}
\lambda^{(0)}_{\rm i}= r_{\rm i} \cos(\theta)~+~ i~r_{\rm i} \sin(\theta), 
\end{eqnarray}

\noindent where, $i^{2}=-1$, $\theta$ is a random number(in radians) in the range $\theta \in [0,~2\pi]$ and $i\in [1,~2,~...,~n]$. These will be initial guesses that begin the iteration process.

The improved version takes less time than the Guggenheimer's version, since in our method, we only calculate U and V once by simply considering the maximum value of the coefficients (given by Eq.  \ref{eqqq}).

Table~\ref{tab:iter} gives the number of iterations that are needed for finding all the roots of a polynomial simultaneously. We used polynomials of third order to three-hundred in the form of equation~\ref{eq:poly} with randomly chosen coefficients. On average it took 2 or 3 iterations to find the roots of a third degree polynomial. Whereas, for a polynomial with the degree of three-hundred, it took about 43 iterations.

\indent Finally, we note that this initial roots algorithm can be used for any root-finding method that requires the input of some initial guess.
\begin{figure}
	\centering
	\includegraphics[width=0.49\textwidth]{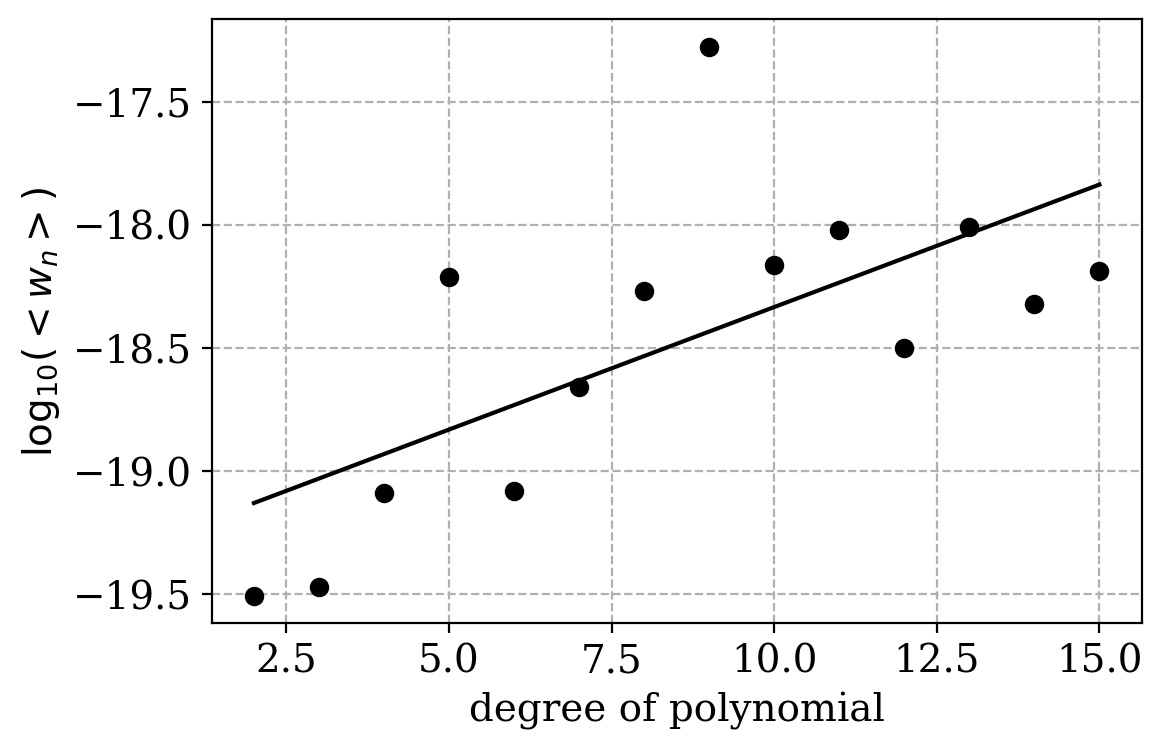}
	\includegraphics[width=0.49\textwidth]{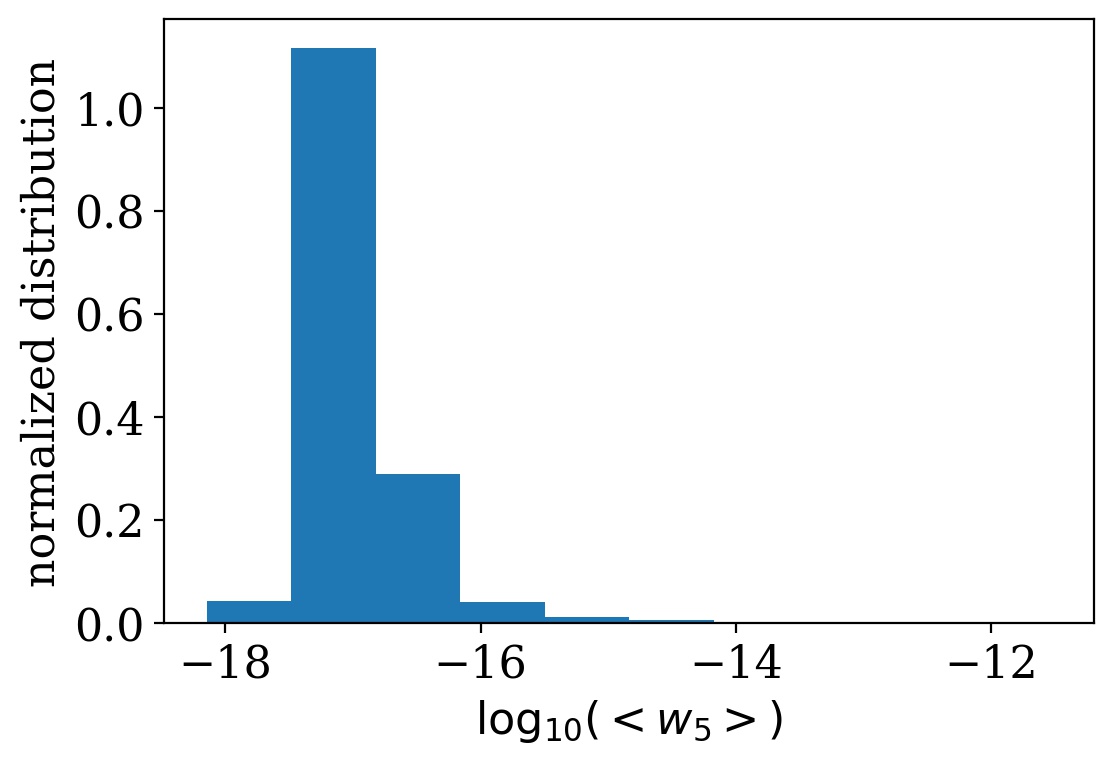}
\caption{Top panel:  The average errors in the roots, $\left< w_{n}\right>$, in Logarithmic scale versus the polynomial's degree $n$. They are calculated with the improved AE algorithm, introduced in subsection \ref{sub1}. Bottom panel: The normalized distribution of errors in roots for polynomials of fifth degree which are solved based on the improved AE algorithm.}
\label{fig:error}
\end{figure}

\begin{table*} 
	\centering
	\begin{tabular}{cccccccccccccc}
		\hline
		\textbf{Degree}: & 3 & 4 & 5 & 6 & 7 & 8 & 9 & 10 & 11 & 12 & 13 & 14 & 15 \\
		\hline
		\textbf{Newton}: & 0.00173 & 0.00226 & 0.00356 & 0.00523 & 0.00772 & 0.00943 & 0.01034 & 0.01081 & 0.01092 & 0.01163 & 0.01317 & 0.01392 & 0.01581 \\
		\textbf{Laguerre}: & 0.00064 & 0.00079 & 0.00135 & 0.00193 & 0.00286 & 0.00347 & 0.00381 & 0.00399 & 0.00403 & 0.00429 & 0.00483 & 0.00519 & 0.00584 \\
		\textbf{AE}: & 0.00018 & 0.00023 & 0.00027 & 0.00054 & 0.00058 & 0.00067 &  0.00072 & 0.00108 & 0.00115 & 0.00127 & 0.00148 & 0.00261 & 0.00314\\
		\hline
	\end{tabular}
	\caption{This table gives the execution time (in seconds) for Newton's, Laguerre's and Aberth-Ehrlich (AE) methods based on the degree of polynomial.}
\label{tab:tim}
\end{table*}
\subsection{Error and stopping criteria}\label{errorstop}

The second part of equation~\ref{eq:Aberth} indicates the value of offset from the exact zeros of polynomial, and we can do the iteration until this offset gets smaller than a desired value, $\epsilon$. Now, The stopping criteria can be defined as:

\begin{eqnarray}
w_{n}=\abs{\frac{\alpha_i^{(k)}}{1-\alpha_i^{(k)}\beta_i^{(k)}}} <\epsilon.
\end{eqnarray}

\noindent So, we can control how accurate the final results are going to be, simply by controlling the value of $\epsilon$. However, the highest achievable accuracy in the finding the roots depends on the polynomial's degree $n$ in addition to the amount of $\epsilon$. Top panel of Figure~\ref{fig:error} demonstrates error for polynomials of second to fifteenth degree with random complex coefficients on a logarithmic scale. The errors were measured by calculating average of absolute values of differences between found roots and the actual zeros of the polynomial, i.e., $\left<w_{n}\right>$, which has a linear behavior in the logarithmic scale, as:
\begin{eqnarray}
\log _{10}( \left<w_{n}\right> ) \simeq -19.3 + 0.1~n,
\end{eqnarray}
where n is polynomial's degree. This equation changes for different values of $\epsilon$. For this plot, we set $\epsilon =10^{-16}$ (the basic accuracy of float numbers in Python programming). Since we are dealing with random polynomials in the process of calculating the error, there may exist some outliers that do not act in a linear way. This is due to the fact that some polynomial may not converge to their actual zeros in the stopping criteria that we define.

Bottom panel of Figure \ref{fig:error} shows the normalized distribution of error for one thousand random complex fifth degree polynomials, similar to the case of binary lens Equations, which is of interest to us. It is apparent that for these series of polynomials, error mostly falls between $10^{-18}$ and $10^{-16}$.

\indent It is known that some root-finding algorithms behave poorly in the cases of degenerate polynomials. A polynomial is called degenerate if it has two distinct roots whose quotient is a root of unity. For example, $(x-1)^2$ is a degenerate polynomial; since its degree is 2, we expect that it should have two roots, but it only has one root, which is 1. In the case of AE, there is no difficulty with finding the roots of a degenerate polynomial. Because, in this algorithm, we first consider n distinct initial guesses, see \ref{Initial_guess}, and then compute the roots with iteration. If a degenerate polynomial has one root, i.e. $\lambda_*^k$ , the initial guesses for this root converge to two points that exist in the following interval:
\begin{eqnarray}
[\lambda_*^k-\epsilon,\lambda_*^k+\epsilon],
\end{eqnarray}
Where $\epsilon$ is the error of computation.
\subsection{Execution Time}
One of the advantages of AE is its short execution time compared to other root-finding algorithms. In table~\ref{tab:tim} we report execution times of this method compared to Newton's and Laguerre's iterations. We chose polynomials with random coefficients varying from third to fifteenth degree and after measuring the time it took for the completion of iterations that produced the roots, we calculated the average execution times. we repeated this one hundred times for each degree. The average time for solving the polynomials with the degrees $3,~4, ...,~15$ for Aberth-Eherlich  was 0.01392 seconds, whereas, it took an average of 0.11653 seconds for Newton's and 0.04302 seconds for Laguerre's method to find the roots. So, AE is about 8 to 9 times faster than Newton's and about 3 to 4 times faster than Laguerre's. Both Newton's and Laguerre's algorithms which are used here for comparing the execution times are available at the Github address \url{https://github.com/AstroFatheddin}. For implementing the laguerre's method, we used the algorithm discussed in \citet{Press2003}. 

For the binary lens equation, which is a fifth degree polynomial with six coefficients, we compared our method with the root-finding algorithm introduced by \citet{SkowronGould2012} (SG). We used ten lens equations with random characteristics and it took about 0.00015 seconds for AE and 0.00029 seconds for SG on average to find all the roots. So, AE is about 1.9 times faster than SG. We note that the time intervals are calculated with the Python module \texttt{time}. 
 
\subsection{Further implementations for Binary Microlensing} 

One way we can improve our method for the lens equation is eliminating the repetition of initial guess process. Instead of using the algorithm stated in \ref{Initial_guess} for every iteration, we only use it once and use the roots that are found in one iteration as initial estimations for the next iteration. This is a logical consideration since the lensing process is dynamic and the source is moving in a path which makes roots, which are the positions of produced images, mostly close to each other. Since we are avoiding the repetition of the initial guess routine, this makes the whole process about 10 times faster without changing the accuracy of the root-finding algorithm.

\indent  Here we also need to note that for a polynomial of fifth degree there exists five distinct roots that are found. But when a source is outside of caustics, only three images are produced. So, two or three of the roots are going to be very close to each other, but the rest are well distinct. In order to check whether a found root is also an answer for the lens equation, we need a fail-safe routine that checks all the found roots in the original lens equation~\ref{eq:lenseq1} and then we can consider the complex roots as a position of an image. But, when the source moves inside caustics, five images are formed and all the roots that are found are image positions.  
   
\section{Conclusions}
Currently, gravitational microlensing observations are done with three observing survey groups, i.e., OGLE, MOA and KMTNet. They totally report more than $3000$ microlensing events annually and this number even increases in near future. An important class of microlensing events are binary and planetary ones, which leads us to discover exoplanets. For better modeling of these binary events, a fast algorithm for solving the binary lens equation (as given by Eq.  \ref{eq:lenseq}) is necessary. Nowadays, all microlensing modelers use Newton's, Laguerre's iterations or their combination (as proposed by \citet{SkowronGould2012}) to find the image positions of point-like source star or points over the source's environment for finite source stars in binary lensing configurations. Hence, the speed of producing binary microlensing lightcurves is limited to the speed of solving the binary lens equation.

\indent In this work, we have introduced an algorithm for solving the binary lens equation based on the Abert-Eldrich method. Additionally, we offered a new method to pick out initial guess solutions inspired by the Guggenheimer's method. For a lens equation (a complex polynomial of fifth degree), our method is faster than the root-finding algorithm introduced by Skowron \& Gould averagely by $1.9$. Hence, our algorithm will improve all binary microlensing modeling methods that use SG to produce light curves by a factor of about $1.9$ to $2.0$.

\indent We also note that in the case of image-centered ray-shooting method \citep{Bennett2009}, where an initial inversion of the lens equation is needed, implementing the AE algorithm can be beneficial.
\section*{acknowledgements}
We thank R. Niyazi for helpful discussions on improving the python implementations of our algorithm. We also thank the referee for his/her useful comments and suggestions.
\section*{Data Availability}
An open source Python code based on our root-finding algorithms discussed in section~\ref{sec:Aberth} is available at:\\ \url{https://github.com/AstroFatheddin/FS-Roots}

Other root-finding algorithms that were mentioned in the text are also available at: \\ \url{https://github.com/AstroFatheddin/Roots}

Our method of finding the initial roots of a polynomial, which we discussed in subsection~\ref{sub1}, can be found at:\\ \url{https://github.com/AstroFatheddin/init}.
\bibliographystyle{mnras}
\bibliography{references} 

\begin{thebibliography}{}
\makeatletter
\relax
\def\mn@urlcharsother{\let\do\@makeother \do\$\do\&\do\#\do\^\do\_\do\%\do\~}
\def\mn@doi{\begingroup\mn@urlcharsother \@ifnextchar [ {\mn@doi@}
  {\mn@doi@[]}}
\def\mn@doi@[#1]#2{\def\@tempa{#1}\ifx\@tempa\@empty \href
  {http://dx.doi.org/#2} {doi:#2}\else \href {http://dx.doi.org/#2} {#1}\fi
  \endgroup}
\def\mn@eprint#1#2{\mn@eprint@#1:#2::\@nil}
\def\mn@eprint@arXiv#1{\href {http://arxiv.org/abs/#1} {{\tt arXiv:#1}}}
\def\mn@eprint@dblp#1{\href {http://dblp.uni-trier.de/rec/bibtex/#1.xml}
  {dblp:#1}}
\def\mn@eprint@#1:#2:#3:#4\@nil{\def\@tempa {#1}\def\@tempb {#2}\def\@tempc
  {#3}\ifx \@tempc \@empty \let \@tempc \@tempb \let \@tempb \@tempa \fi \ifx
  \@tempb \@empty \def\@tempb {arXiv}\fi \@ifundefined
  {mn@eprint@\@tempb}{\@tempb:\@tempc}{\expandafter \expandafter \csname
  mn@eprint@\@tempb\endcsname \expandafter{\@tempc}}}

\bibitem[\protect\citeauthoryear{Aberth}{Aberth}{1973}]{Aberth1973}
Aberth O.,  1973, Mathematics of Computation, 27, 339

\bibitem[\protect\citeauthoryear{{Bachelet}, {Norbury}, {Bozza}  \&
  {Street}}{{Bachelet} et~al.}{2017}]{pyLima}
{Bachelet} E.,  {Norbury} M.,  {Bozza} V.,   {Street} R.,  2017, \mn@doi [\aj]
  {10.3847/1538-3881/aa911c}, \href
  {https://ui.adsabs.harvard.edu/abs/2017AJ....154..203B} {154, 203}

\bibitem[\protect\citeauthoryear{{Bennett}}{{Bennett}}{2008}]{Bennett2008}
{Bennett} D.~P.,  2008, in {Mason} J.~W.,  ed., , Exoplanets.
p.~47, \mn@doi{10.1007/978-3-540-74008-7\_3}

\bibitem[\protect\citeauthoryear{{Bennett}}{{Bennett}}{2010}]{Bennett2009}
{Bennett} D.~P.,  2010, \mn@doi [\apj] {10.1088/0004-637X/716/2/1408}, \href
  {https://ui.adsabs.harvard.edu/abs/2010ApJ...716.1408B} {716, 1408}

\bibitem[\protect\citeauthoryear{Bennett \& Rhie}{Bennett \&
  Rhie}{1996}]{Bennett1996}
Bennett D.~P.,  Rhie S.~H.,  1996, \mn@doi [The Astrophysical Journal]
  {10.1086/178096}, 472, 660

\bibitem[\protect\citeauthoryear{{Bozza}}{{Bozza}}{2010}]{Bozza2010}
{Bozza} V.,  2010, \mn@doi [\mnras] {10.1111/j.1365-2966.2010.17265.x}, \href
  {https://ui.adsabs.harvard.edu/abs/2010MNRAS.408.2188B} {408, 2188}

\bibitem[\protect\citeauthoryear{{Bozza}, {Bachelet}, {Bartoli{\'c}}, {Heintz},
  {Hoag}  \& {Hundertmark}}{{Bozza} et~al.}{2018}]{2018Bozza}
{Bozza} V.,  {Bachelet} E.,  {Bartoli{\'c}} F.,  {Heintz} T.~M.,  {Hoag} A.~R.,
    {Hundertmark} M.,  2018, \mn@doi [\mnras] {10.1093/mnras/sty1791}, \href
  {https://ui.adsabs.harvard.edu/abs/2018MNRAS.479.5157B} {479, 5157}

\bibitem[\protect\citeauthoryear{{Dominik}}{{Dominik}}{1998}]{1998Dominik}
{Dominik} M.,  1998, \aap, \href
  {https://ui.adsabs.harvard.edu/abs/1998A&A...333L..79D} {333, L79}

\bibitem[\protect\citeauthoryear{Ehrlich}{Ehrlich}{1967}]{Ehrlich1967}
Ehrlich L.~W.,  1967, \mn@doi [Commun. ACM] {10.1145/363067.363115}, 10,
  107–108

\bibitem[\protect\citeauthoryear{{Erdl} \& {Schneider}}{{Erdl} \&
  {Schneider}}{1993}]{1993Erdlschneider}
{Erdl} H.,  {Schneider} P.,  1993, \aap, \href
  {https://ui.adsabs.harvard.edu/abs/1993A&A...268..453E} {268, 453}

\bibitem[\protect\citeauthoryear{{Gaudi}}{{Gaudi}}{2012}]{gaudi2012}
{Gaudi} B.~S.,  2012, \mn@doi [\araa] {10.1146/annurev-astro-081811-125518},
  50, 411

\bibitem[\protect\citeauthoryear{Ghidouche, Sider, Couturier  \&
  Guyeux}{Ghidouche et~al.}{2017}]{Ghidouche2017}
Ghidouche K.,  Sider A.,  Couturier R.,   Guyeux C.,  2017, J. Comput. Sci.,
  18, 46

\bibitem[\protect\citeauthoryear{{Gould} \& {Gaucherel}}{{Gould} \&
  {Gaucherel}}{1997}]{1997Gould}
{Gould} A.,  {Gaucherel} C.,  1997, \mn@doi [\apj] {10.1086/303751}, \href
  {https://ui.adsabs.harvard.edu/abs/1997ApJ...477..580G} {477, 580}

\bibitem[\protect\citeauthoryear{{Gould} \& {Loeb}}{{Gould} \&
  {Loeb}}{1992}]{Gould1992}
{Gould} A.,  {Loeb} A.,  1992, \mn@doi [\apj] {10.1086/171700}, \href
  {https://ui.adsabs.harvard.edu/abs/1992ApJ...396..104G} {396, 104}

\bibitem[\protect\citeauthoryear{Guggenheimer}{Guggenheimer}{1986}]{Guggenheimer1986}
Guggenheimer H.,  1986, BIT Numerical Mathematics, 26, 537

\bibitem[\protect\citeauthoryear{{Kayser}, {Refsdal}  \& {Stabell}}{{Kayser}
  et~al.}{1986}]{1986Kayser}
{Kayser} R.,  {Refsdal} S.,   {Stabell} R.,  1986, \aap, \href
  {https://ui.adsabs.harvard.edu/abs/1986A&A...166...36K} {166, 36}

\bibitem[\protect\citeauthoryear{{Kim} et~al.,}{{Kim}
  et~al.}{2016}]{KMTNet2016}
{Kim} S.-L.,  et~al., 2016, \mn@doi [Journal of Korean Astronomical Society]
  {10.5303/JKAS.2016.49.1.037}, \href
  {https://ui.adsabs.harvard.edu/abs/2016JKAS...49...37K} {49, 37}

\bibitem[\protect\citeauthoryear{{Mao} \& {Paczynski}}{{Mao} \&
  {Paczynski}}{1991}]{1991MAO}
{Mao} S.,  {Paczynski} B.,  1991, \mn@doi [\apjl] {10.1086/186066}, \href
  {https://ui.adsabs.harvard.edu/abs/1991ApJ...374L..37M} {374, L37}

\bibitem[\protect\citeauthoryear{{Mr{\'o}z} et~al.,}{{Mr{\'o}z}
  et~al.}{2019}]{2019Mroz}
{Mr{\'o}z} P.,  et~al., 2019, \mn@doi [\aap] {10.1051/0004-6361/201834557},
  \href {https://ui.adsabs.harvard.edu/abs/2019A&A...622A.201M} {622, A201}

\bibitem[\protect\citeauthoryear{{Penny} et~al.,}{{Penny}
  et~al.}{2013}]{Penny2013}
{Penny} M.~T.,  et~al., 2013, \mn@doi [\mnras] {10.1093/mnras/stt927}, \href
  {https://ui.adsabs.harvard.edu/abs/2013MNRAS.434....2P} {434, 2}

\bibitem[\protect\citeauthoryear{{Penny}, {Gaudi}, {Kerins}, {Rattenbury},
  {Mao}, {Robin}  \& {Calchi Novati}}{{Penny} et~al.}{2019}]{Penny2019}
{Penny} M.~T.,  {Gaudi} B.~S.,  {Kerins} E.,  {Rattenbury} N.~J.,  {Mao} S.,
  {Robin} A.~C.,   {Calchi Novati} S.,  2019, \mn@doi [\apjs]
  {10.3847/1538-4365/aafb69}, \href
  {https://ui.adsabs.harvard.edu/abs/2019ApJS..241....3P} {241, 3}

\bibitem[\protect\citeauthoryear{{Poleski} \& {Yee}}{{Poleski} \&
  {Yee}}{2019}]{MuLensModel}
{Poleski} R.,  {Yee} J.~C.,  2019, \mn@doi [Astronomy and Computing]
  {10.1016/j.ascom.2018.11.001}, \href
  {https://ui.adsabs.harvard.edu/abs/2019A&C....26...35P} {26, 35}

\bibitem[\protect\citeauthoryear{Press, Teukolsky, Vettering  \&
  Flannery}{Press et~al.}{2003}]{Press2003}
Press W.~H.,  Teukolsky S.~A.,  Vettering W.~T.,   Flannery B.~P.,  2003,
  \mn@doi [European Journal of Physics] {10.1088/0143-0807/24/3/701}, 24, 329

\bibitem[\protect\citeauthoryear{{Sahu} et~al.,}{{Sahu}
  et~al.}{2022}]{2022Sahu}
{Sahu} K.~C.,  et~al., 2022, arXiv e-prints, \href
  {https://ui.adsabs.harvard.edu/abs/2022arXiv220113296S} {p. arXiv:2201.13296}

\bibitem[\protect\citeauthoryear{{Sajadian}}{{Sajadian}}{2021a}]{2021sajadianb}
{Sajadian} S.,  2021a, \mn@doi [\mnras] {10.1093/mnras/stab1907}, \href
  {https://ui.adsabs.harvard.edu/abs/2021MNRAS.506.3615S} {506, 3615}

\bibitem[\protect\citeauthoryear{{Sajadian}}{{Sajadian}}{2021b}]{2021sajadian}
{Sajadian} S.,  2021b, \mn@doi [\mnras] {10.1093/mnras/stab2942}, \href
  {https://ui.adsabs.harvard.edu/abs/2021MNRAS.508.5991S} {508, 5991}

\bibitem[\protect\citeauthoryear{{Sajadian} \& {Rahvar}}{{Sajadian} \&
  {Rahvar}}{2010}]{2010sajadian}
{Sajadian} S.,  {Rahvar} S.,  2010, \mn@doi [\mnras]
  {10.1111/j.1365-2966.2010.16901.x}, \href
  {https://ui.adsabs.harvard.edu/abs/2010MNRAS.407..373S} {407, 373}

\bibitem[\protect\citeauthoryear{{Sako} et~al.,}{{Sako}
  et~al.}{2008}]{MOA_gourp}
{Sako} T.,  et~al., 2008, \mn@doi [Experimental Astronomy]
  {10.1007/s10686-007-9082-5}, \href
  {https://ui.adsabs.harvard.edu/abs/2008ExA....22...51S} {22, 51}

\bibitem[\protect\citeauthoryear{{Schneider} \& {Weiss}}{{Schneider} \&
  {Weiss}}{1986}]{1986Schneider}
{Schneider} P.,  {Weiss} A.,  1986, \aap, \href
  {https://ui.adsabs.harvard.edu/abs/1986A&A...164..237S} {164, 237}

\bibitem[\protect\citeauthoryear{{Schneider}, {Ehlers}  \& {Falco}}{{Schneider}
  et~al.}{1992}]{1992schneider}
{Schneider} P.,  {Ehlers} J.,   {Falco} E.~E.,  1992, {Gravitational Lenses},
  \mn@doi{10.1007/978-3-662-03758-4.
}

\bibitem[\protect\citeauthoryear{{Simpson}}{{Simpson}}{1740}]{Simpson1740}
{Simpson} T.,  1740, Woodfall, 6, 81

\bibitem[\protect\citeauthoryear{{Skowron} \& {Gould}}{{Skowron} \&
  {Gould}}{2012}]{SkowronGould2012}
{Skowron} J.,  {Gould} A.,  2012, arXiv e-prints, \href
  {https://ui.adsabs.harvard.edu/abs/2012arXiv1203.1034S} {p. arXiv:1203.1034}

\bibitem[\protect\citeauthoryear{{Udalski}, {Szyma{\'n}ski}  \&
  {Szyma{\'n}ski}}{{Udalski} et~al.}{2015}]{OGLE_IV}
{Udalski} A.,  {Szyma{\'n}ski} M.~K.,   {Szyma{\'n}ski} G.,  2015, \actaa,
  \href {https://ui.adsabs.harvard.edu/abs/2015AcA....65....1U} {65, 1}

\bibitem[\protect\citeauthoryear{{Wambsganss}}{{Wambsganss}}{1990}]{1990Wambsganss}
{Wambsganss} J.,  1990, PhD thesis, -

\bibitem[\protect\citeauthoryear{{Wambsganss}}{{Wambsganss}}{1999}]{1999Wambsganss}
{Wambsganss} J.,  1999, Journal of Computational and Applied Mathematics, \href
  {https://ui.adsabs.harvard.edu/abs/1999JCoAM.109..353W} {109, 353}

\bibitem[\protect\citeauthoryear{{Witt}}{{Witt}}{1990}]{Witt1990}
{Witt} H.~J.,  1990, \aap, 236, 311

\bibitem[\protect\citeauthoryear{{Witt} \& {Mao}}{{Witt} \&
  {Mao}}{1994}]{1994wittmoa}
{Witt} H.~J.,  {Mao} S.,  1994, \mn@doi [\apj] {10.1086/174426}, \href
  {https://ui.adsabs.harvard.edu/abs/1994ApJ...430..505W} {430, 505}

\bibitem[\protect\citeauthoryear{{Witt} \& {Mao}}{{Witt} \&
  {Mao}}{1995}]{WittMao1995}
{Witt} H.~J.,  {Mao} S.,  1995, \mn@doi [\apjl] {10.1086/309566}, \href
  {https://ui.adsabs.harvard.edu/abs/1995ApJ...447L.105W} {447, L105}

\makeatother
\end{thebibliography}
\end{document}